\documentclass[a4paper]{jpconf}
\usepackage{amsmath,amsfonts,amssymb,bm}
\usepackage{dsfont}
\usepackage{graphicx}
\usepackage{color}
\usepackage{soul}
\usepackage[comma,square,sort,compress,numbers]{natbib}

\usepackage[mathscr,scaled=1.15]{urwchancal}
\DeclareFontFamily{OT1}{pzc}{}
\DeclareFontShape{OT1}{pzc}{m}{it}%
{<-> s * [1.15] pzcmi7t}{}
\DeclareMathAlphabet{\mathpzc}{OT1}{pzc}{m}{it}

\definecolor{purple}{rgb}{0.5,0,0.5}
\definecolor{blue}{rgb}{0.0,0,0.9}
\definecolor{prdblue}{rgb}{0.133,0.118,0.498}
\usepackage[colorlinks=true, pdfstartview=FitV, linkcolor=prdblue, citecolor= prdblue, urlcolor=prdblue]{hyperref}

\begin{document}
\title{$\,$\\[-6ex]\hspace*{\fill}{\sf\small{\emph{Preprint no}. NJU-INP 005/19}}\\[1ex]Insights into the Origin of Mass}

\author{Craig D. Roberts}

\address{School of Physics, Nanjing University, Nanjing, Jiangsu 210093, China}
\address{Institute for Nonperturbative Physics, Nanjing University, Nanjing, Jiangsu 210093, China}

\ead{cdroberts@nju.edu.cn}

\begin{abstract}
Atomic nuclei are the core of everything we can see.  At the first level of approximation, their atomic weights are simply the sum of the masses of all the nucleons they contain.  Each nucleon has a mass $m_N \approx 1\,$GeV, \emph{i.e}.\ approximately 2000-times the electron mass. The Higgs boson produces the latter, but what produces the nucleon mass?  This is the crux: the vast bulk of the mass of a nucleon is lodged with the energy needed to hold quarks together inside it; and that is supposed to be explained by quantum chromodynamics (QCD), the strong-interaction piece within the Standard Model.  This contribution canvasses the potential for a coherent effort in QCD phenomenology and theory, coupled with experiments at existing and planned facilities, to reveal the origin and distribution of mass by focusing on the properties of the strong-interaction Nambu-Goldstone modes.  Key experiments are approved at JLab 12; planned with COMPASS++/AMBER at CERN; and could deliver far-reaching insights by exploiting the unique capabilities foreseen at an electron ion collider.
\end{abstract}


\section{In the Beginning}
The Higgs boson (or something very like it) has been discovered at the Large Hadron Collider (LHC) \cite{Aad:2012tfa, Chatrchyan:2012xdj}.  As a consequence, the 2013 Nobel Prize in physics was awarded to F.~Englert and P.~Higgs \cite{Englert:2014zpa, Higgs:2014aqa}: ``for the theoretical discovery of a mechanism that contributes to our understanding of the origin of mass of subatomic particles \ldots''.  This achievement has consummated the Standard Model of Particle Physics.  Notwithstanding that, there are empirical features of Nature that lie beyond a Standard Model explanation, \emph{e.g}.\ dark matter and dark energy.  However, these are extraterrestrial signatures; unambiguous signals of physics beyond the Standard Model otherwise continue to be elusive.

The hunt goes on; and yet, the race to find physics beyond the Standard Model has left some of the most fundamental aspects of Nature unexplained.  Everything we see and use is built from atoms and molecules.  Their energy levels and electromagnetic properties are readily understood, using quantum mechanics augmented by quantum electrodynamics (QED) at higher energies.  Within every atom, however, lies a compact nucleus, comprised of neutrons and protons; and the structure and arrangements of all these things is supposed to be described by quantum chromodynamics (QCD), the strong interaction part of the Standard Model.  The status of QCD and its predictions was highlighted in the Nobel Prize acceptance speech given by H.\,D.~Politzer; in particular, the following remarks \cite{Politzer:2005kc}:
``\emph{The establishment by the mid-1970's of QCD as the correct theory of the strong interactions completed what is now known prosaically as the Standard Model.  It offers a description of all known fundamental physics except for gravity, and gravity is something that has no discernible effect when particles are studied a few at a time.  However, the situation is a bit like the way that the Navier-Stokes equation accounts for the flow of water.  The equations are at some level obviously correct, but there are only a few, limited circumstances in which their consequences can be worked out in any detail.''}

The issue here is that the Higgs is often said to give mass to everything, but\footnote{\href{https://www.theguardian.com/science/2011/dec/13/higgs-boson-seminar-god-particle}
{theguardian.com/science/2011/dec/13/higgs-boson-seminar-god-particle}}: ``\emph{That is wrong.  The Higgs field only gives mass to some very simple particles.  The field accounts for only one or two percent of the mass of more complex things like atoms, molecules and everyday objects, from your mobile phone to your pet llama. The vast majority of mass comes from the energy needed to hold quarks together inside atoms}.''  These remarks highlight QCD, the quantum field theory formulated in four spacetime dimensions which defines what is arguably the most important piece of the Standard Model.  QCD is supposed to describe all of nuclear physics through interactions between quarks (matter fields) mediated by gluons (gauge bosons).  Yet, fifty years after the discovery of quarks, science is only just beginning to grasp how QCD moulds pions, nucleons, \emph{etc}.; and it is far from understanding how QCD produces nuclei.

The natural mass-scale for nuclear physics is characterised by the proton mass:
$m_p \approx 1\,{\rm GeV} \approx 2000\,m_e$, 
where $m_e$ is the electron mass.  In the Standard Model, $m_e$ is rightly attributed to the Higgs boson; but what is the source of the enormous enhancement required to produce $m_p$?  Followed logically to its origin, this question leads to an appreciation that the existence of our Universe depends critically on, \emph{inter alia}, the following empirical facts:
(\emph{i}) the proton is massive, \emph{i.e}.\ the mass-scale for strong interactions is vastly different to that of electromagnetism;
(\emph{ii}) the proton is absolutely stable, despite being a composite object constituted from three valence-quarks;
and (\emph{iii}) the pion, responsible for long-range interactions between nucleons is unnaturally light (not massless), possessing a lepton-like mass despite being a strongly interacting composite object built from a valence-quark and valence antiquark.  (Discovered in 1936 \cite{PhysRev.51.884}, the $\mu$-lepton was initially mistaken for the pion.  The pion was actually found a decade later \cite{Lattes:1947mw}.)

The Lagrangian of chromodynamics is simple: ${\mathpzc L}_{\rm QCD}$ can be expressed in one short line with two supplementary definitions.  Comparing with QED, the \emph{solitary} difference is the term describing gluon self-interactions, which appears because chromodynamics is built upon the non-Abelian SU$(3)$ gauge-group.  Nevertheless, somehow, ${\mathpzc L}_{\rm QCD}$ -- this one line, along with two definitions -- is responsible for the origin, mass, and size of almost all visible matter in the Universe.  QCD is thus, quite possibly, the most remarkable fundamental theory ever invented.

The only apparent energy-scales in ${\mathpzc L}_{\rm QCD}$ are the current-quark masses, generated by the Higgs boson; but focusing on the quark flavours that define nucleons, \emph{i.e}.\ $u$ (up) and $d$ (down) quarks, this scale is more-than two orders-of-magnitude smaller than $m_p$.  No amount of ``staring'' at ${\mathpzc L}_{\rm QCD}$ can reveal the source of that enormous amount of ``missing mass''; yet, it must be there.  This is a stark contrast to QED wherein, \emph{e.g}.\ the scale in the spectrum of the hydrogen atom is set by $m_e$, a prominent feature of ${\mathpzc L}_{\rm QED}$ that is generated by the Higgs boson.

Treated as a classical model, chromodynamics is a non-Abelian local gauge theory.  Formulated in four dimensions, such theories possess no mass-scale in the absence of Lagrangian masses for the matter fields.  (This circumstance defines the chiral limit.) There is no dynamics in a scale-invariant theory, only kinematics: the theory looks the same at all length-scales; hence, bound states are impossible and, accordingly, our Universe cannot exist.
A spontaneous breaking of symmetry, as realised via the Higgs mechanism, does not solve this problem: the masses of the neutron and proton, the kernels of all visible matter, are $\sim 100$-times larger than the Higgs-generated current-masses of the $u$- and $d$-quarks, the main building blocks of nucleons.
Consequently, the questions of how does a mass-scale appear and why does it have the value we observe are inseparable from the question: ``How did the Universe come into being?''

In a scale invariant theory, Poincar\'e invariance entails that the energy-momentum tensor is traceless: $T_{\mu\mu} \equiv 0$.  However, classical chromodynamics is not a meaningful physical framework.  It must be quantised; and in completing that operation, the regularisation and renormalisation of ultraviolet divergences introduces a mass-scale.  This is ``dimensional transmutation'': all quantities, including the field operators themselves, become dependent on a mass-scale; and, consequently, one encounters the chiral-limit ``trace anomaly'':
\begin{equation}
\label{SIQCD}
T_{\mu\mu} = \beta(\alpha(\zeta))  \tfrac{1}{4} G^{a}_{\mu\nu}G^{a}_{\mu\nu} =: \Theta_0 \,,
\end{equation}
where $\beta(\alpha(\zeta))$ is QCD's $\beta$-function, $\alpha(\zeta)$ is the associated running-coupling, and $\zeta$ is the renormalisation scale.  Eq.\,\eqref{SIQCD} indicates that a mass-scale related to the resolving power of a given measurement is introduced via quantisation, \emph{viz}.\ the scale \emph{emerges} as an integral part of the theory's quantum definition.

Knowing that a trace anomaly exists does not deliver much: it only indicates that there is a mass-scale.  The crucial issue is whether one can understand the magnitude of that scale.
One can certainly measure its size, for consider the energy-momentum tensor in the proton:
\begin{equation}
\label{EPTproton}
\langle p(P) | T_{\mu\nu} | p(P) \rangle = - P_\mu P_\nu\,,
\end{equation}
where the right-hand-side follows from the equations-of-motion for a one-particle proton state.  Now, in the chiral limit,
\begin{align}
\label{anomalyproton}
\langle p(P) | T_{\mu\mu} | p(P) \rangle  = - P^2 & = m_p^2 = \langle p(P) |  \Theta_0 | p(P) \rangle\,;
\end{align}
namely, there is a clear sense in which it is possible to conclude that the entirety of the proton mass is produced by gluons.  The trace anomaly is measurably large; and that property must logically owe to gluon self-interactions, which are also responsible for asymptotic freedom.  This is what is meant by the quote above: ``\emph{The vast majority of mass comes from the energy needed to hold quarks together inside atoms}.''

There is also a flip-side, which can be exposed by replacing the proton by the pion in Eq.\,\eqref{EPTproton}:
\begin{equation}
\label{EPTpion}
\langle \pi(q) | T_{\mu\nu} | \pi(q) \rangle = - q_\mu q_\nu\,
\stackrel{\rm chiral\,limit}{\Rightarrow} \langle \pi(q) |  \Theta_0 | \pi (q) \rangle = 0
\end{equation}
because the pion is a massless Nambu-Goldstone (NG) mode.  Equation\,\eqref{EPTpion} could mean that the scale anomaly vanishes trivially in the pion state, \emph{viz}.\ that gluons and their self-interactions have no impact within a pion because each term in the practical computation of the operator expectation value vanishes when evaluated in the pion. However, that is a difficult way to achieve Eq.\,\eqref{EPTpion}.  It is easier to imagine that Eq.\,\eqref{EPTpion} owes to cancellations between different operator-component contributions.  Of course, such precise cancellation should not be an accident.  It could only arise naturally because of some symmetry and/or symmetry-breaking pattern.

Eqs.\,\eqref{anomalyproton} and \eqref{EPTpion} present a quandary, highlighting that no understanding of the source of the proton's mass can be complete unless it simultaneously explains the meaning of Eq.\,\eqref{EPTpion}.  Moreover, any discussion of confinement, fundamental to the proton's absolute stability, is impossible before this conundrum is resolved.  The explanation of these features of Nature lies in the dynamics responsible for the emergence of $m_p$ as the natural nuclear-physics mass-scale; and one of the most important goals in science is to elucidate all consequences of this dynamics.

\section{Gluons are Massive and so the Coupling Saturates}
\label{SecGluonMass}
Gluons must be the key; after all, their self interactions separate QCD from QED.  Gluons are supposed to be massless.  This is true perturbatively; but it is a feature that is not preserved nonperturbatively.  Beginning with a pioneering effort almost forty years ago \cite{Cornwall:1981zr}, continuum and lattice studies of QCD's gauge sector have been increasing in sophistication and reliability; and today it is known that the gluon propagator saturates at infrared momenta \cite{Boucaud:2011ug, Aguilar:2015bud, Gao:2017uox, Cyrol:2017ewj}:
\begin{equation}
\label{eqGluonMass}
\Delta(k^2\simeq 0) = 1/m_g^2.
\end{equation}
Thus, the long-range propagation characteristics of gluons are dramatically affected by their self-interactions.  Importantly, one may associate a renormalisation-group-invariant gluon mass-scale with this effect: $m_0 \approx 0.5\,$GeV\,$\approx m_p/2$, and summarise a large body of work by stating that gluons, although acting as massless degrees-of-freedom on the perturbative domain, actually possess a running mass, whose value at infrared momenta is characterised by $m_0$.

Asymptotic freedom ensures that QCD's ultraviolet behaviour is controllable; but the emergence of a gluon mass reveals a new frontier within the Standard Model because the existence of a running gluon mass, large at infrared momenta, has an impact on all analyses of the bound-state problem.  Furthermore, $m_0>0 $ entails that QCD dynamically generates its own infrared cutoff, so that gluons with wavelengths $\lambda \gtrsim \sigma :=1/m_0 \approx 0.5\,$fm decouple from the strong interaction \cite{Gao:2017uox}, hinting at a dynamical realisation of confinement \cite{Brodsky:2015aia}.

\begin{figure}[t]
\leftline{\includegraphics[width=0.46\textwidth]{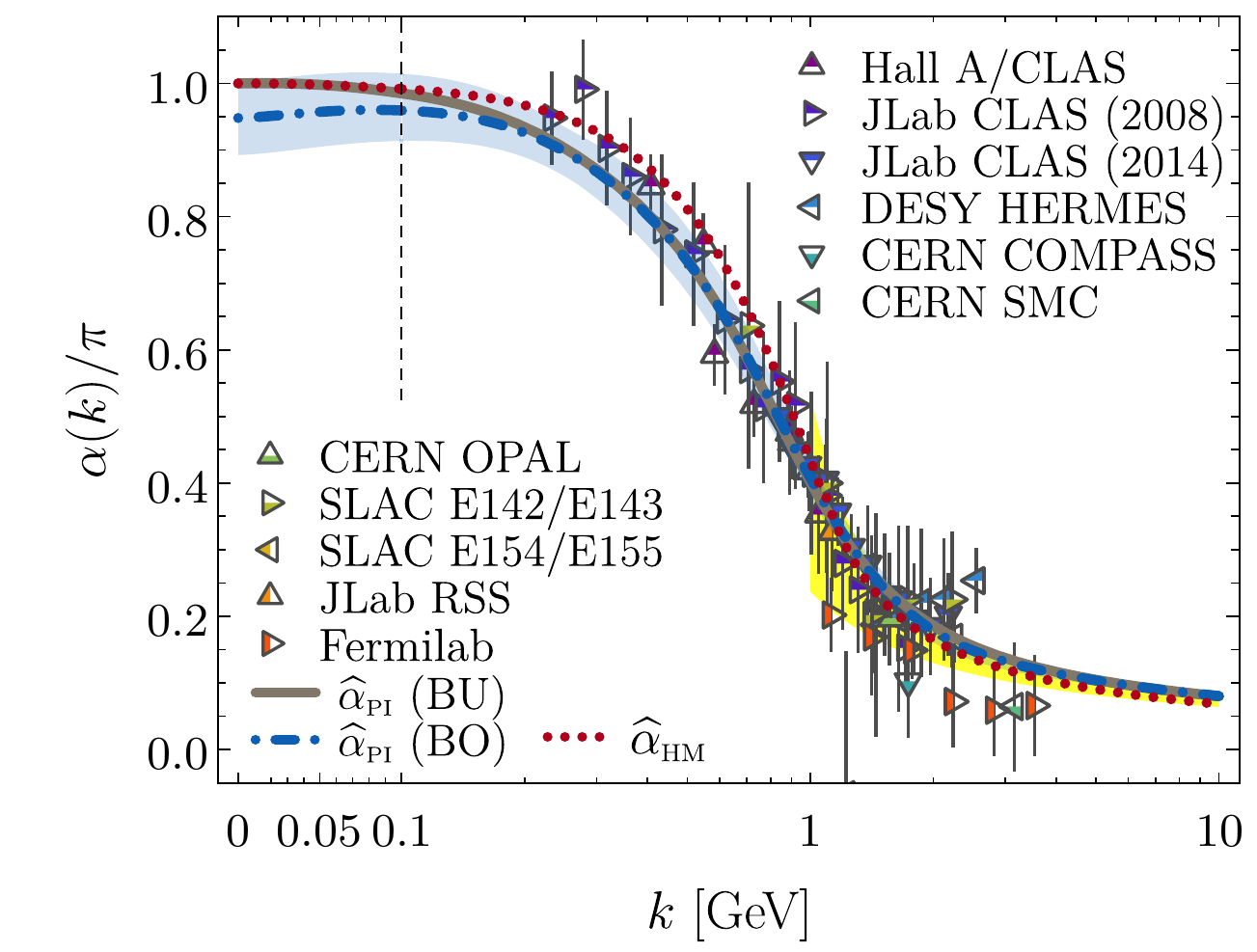}}
\vspace*{-41ex}

\hspace*{0.5\textwidth}\parbox[t]{0.45\textwidth}{\caption{\label{FigwidehatalphaII}\small
Dot-dashed [blue] curve: process-in\-de\-pen\-dent running-coupling $\alpha_{\rm PI}(k^2)$ \cite{Binosi:2016nme}: the shaded [blue] band bracketing this curve combines a 95\% confidence-level window based on existing lQCD results for the gluon two-point function with an error of 10\% in the continuum analysis of relevant ghost-gluon dynamics.
Solid [black] curve, updated result \cite{Rodriguez-Quintero:2018wma}.
Also depicted, world's data on the process-\underline{dependent} coupling $\alpha_{g_1}$, defined via the Bjorken sum rule, the sources of which are listed elsewhere \cite{Binosi:2016nme}.
The shaded [yellow] band on $k>1\,$GeV represents $\alpha_{g_1}$ obtained from the Bjorken sum by using QCD evolution to extrapolate high-$k^2$ data into the depicted region \cite{Deur:2008rf}; and, for additional context, the dashed [red] curve is the effective charge obtained in a light-front holographic model \cite{Deur:2016tte}.
%
}}
\end{figure}

There are many other consequences of the intricate nonperturbative nature of QCD's gauge-sector dynamics.  Important amongst them is the generation of a process-independent running coupling, $\alpha_{\rm PI}(k^2)$ \cite{Binosi:2016nme, Rodriguez-Quintero:2018wma}, the solid [black] curve in Fig.\,\ref{FigwidehatalphaII}.   This charge is an analogue of the Gell-Mann--Low effective coupling in QED because it is completely determined by the gauge-boson propagator.  The result in Fig.\,\ref{FigwidehatalphaII} is a parameter-free prediction, capitalising on continuum analyses of QCD's gauge sector informed by simulations of lattice-QCD (lQCD).

As a unique process-independent effective charge, $\alpha_{\rm PI}$ appears in every one of QCD's dynamical equations of motion, including the quark gap equation, setting the strength of all interactions.  $\alpha_{\rm PI}$ therefore plays a crucial role in understanding the dynamical origin of light-quark masses in the Standard Model even in the absence of a Higgs coupling.

It is worth observing here that Fig.\,\ref{FigwidehatalphaII} shows QCD's effective coupling to be everywhere finite, \emph{viz}.\ there is no Landau pole and the theory likely possesses an infrared-stable fixed point.  These features owe to the dynamical generation of a gluon mass scale.  At the other extreme, asymptotic freedom guarantees that QCD is well-defined at ultraviolet momenta.  QCD is therefore unique amongst known four-dimensional quantum field theories: potentially, it is defined and internally consistent at all energy scales.  As such, it may serve to establish a paradigm for understanding physics beyond the Standard Model using strongly-coupled non-Abelian gauge theories.

\begin{figure}[t]
\leftline{\includegraphics[clip,width=0.43\textwidth]{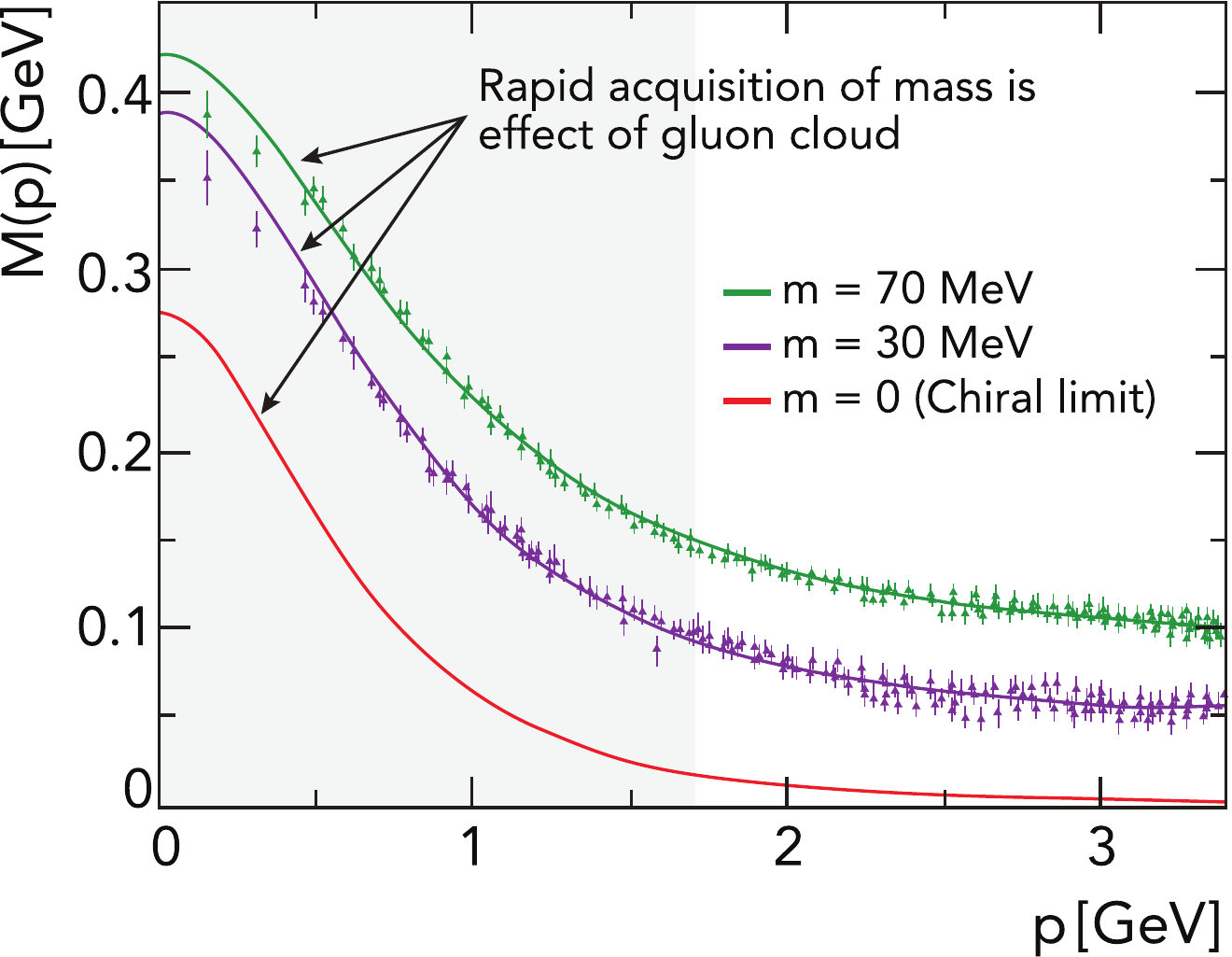}}
\vspace*{-37ex}

\hspace*{0.5\textwidth}\parbox[t]{0.45\textwidth}{\caption{\label{gluoncloud}
\small
Renormalisation-group-invariant dressed-quark mass function, $M(p)$ in Eq.\,\eqref{Spgen}: \emph{solid curves} -- continuum nonperturbative results \cite{Bhagwat:2003vw, Bhagwat:2006tu}; ``data'' -- numerical simulations of lQCD \protect\cite{Bowman:2005vx}.
The size of $M(0)$ is a measure of the magnitude of the QCD scale anomaly in $n=1$-point Schwinger functions \cite{Roberts:2016vyn}.
Experiments on $Q^2\in [0,12]\,$GeV$^2$ at the modern Thomas Jefferson National Accelerator Facility (JLab) will be sensitive to the momentum dependence of $M(p)$ within a domain that is here indicated approximately by the shaded region.
}}\vspace*{2ex}
\end{figure}

The emergence of a gluon mass-scale in the Standard Model drives an enormous array of phenomena.  Crucial amongst them is dynamical chiral symmetry breaking (DCSB), which is most readily apparent in the dressed-quark propagator:
\begin{equation}
\label{Spgen}
S(p) = 1/[i \gamma\cdot p A(p^2) + B(p^2)] = Z(p^2)/[i\gamma\cdot p + M(p^2)]\,.
\end{equation}
$S(p)$ can be calculated in QCD using nonperturbative continuum and lattice techniques.  $M(p^2)$ in Eq.\,\eqref{Spgen} is the dressed-quark mass-function.  Its predicted behaviour is depicted in Fig.\,\ref{gluoncloud}, which reveals that the current-quark of perturbative QCD evolves into a constituent-quark as its momentum becomes smaller.  The constituent-quark mass arises from a cloud of low-momentum gluons attaching themselves to the current-quark.  This is DCSB, the essentially nonperturbative effect that generates a quark \emph{mass} \emph{from nothing}; namely, it occurs even in the absence of a Higgs mechanism.  Evidently, even in this case, when current-masses vanish, quarks acquire a running mass whose value in the infrared is approximately $m_p/3$.  This is the scale required to support the constituent quark model and all its successes.  It follows that DCSB can be identified as the source for more than 98\% of the visible mass in the Universe.

DCSB is very clearly revealed in pion properties.  In fact \cite{Roberts:2016vyn}, the key to understanding Eq.\,\eqref{EPTpion} is a set of Goldberger-Treiman (GT) relations \cite{Qin:2014vya, Binosi:2016rxz}, the best known of which states:
\begin{equation}
\label{GTRE}
m \simeq 0 \; \left| \; f_\pi E_\pi(k;0) = B(k^2)
\right. ,
\end{equation}
where $E_\pi$ is the leading piece of the pion's Bethe-Salpeter amplitude (simply related to its wave function) and $B$ is the dressed-quark's scalar self-energy, Eq.\,\eqref{Spgen}.  Eq.\,\eqref{GTRE} is exact in chiral QCD and expresses the fact that the Nambu-Goldstone theorem is fundamentally an expression of equivalence between the quark one-body problem and the two-body bound-state problem in QCD's flavor-nonsinglet pseudoscalar meson channel. Consequently and enigmatically:\\[0.5ex]
\hspace*{1.8em}\parbox[t]{0.91\textwidth}{
\emph{The properties of the nearly-massless pion are the cleanest expression of the mechanism that is responsible for (almost) all the visible mass in the Universe}.}
\smallskip

\hspace*{-\parindent}Eq.\,\eqref{GTRE} has many corollaries, \emph{e.g}.\ it proves both that the pion exists as a NG mode if, and only if, mass emerges dynamically and that the $\pi$-nucleon coupling is strong if, and only if, the scale of light-quark emergent mass is large.

With the GT relations in hand, one can construct an algebraic proof \cite{Qin:2014vya, Binosi:2016rxz}, that at any and each order in a symmetry-preserving truncation of those equations in quantum field theory necessary to describe a pseudoscalar bound state, there is a precise cancellation between the mass-generating effect of dressing the valence-quark and -antiquark which constitute the system and the attraction generated by the interactions between them, \emph{i.e}.
\begin{align}
M^{\rm dressed}_{\rm quark} + M^{\rm dressed}_{\rm antiquark}+ U^{\rm dressed}_{\rm quark-antiquark\;interaction} \stackrel{\rm chiral\;limit}{\equiv} 0\,.
\label{EasyOne}
\end{align}
This guarantees the ``disappearance'' of the scale anomaly in the chiral-limit pion.
An analogy with quantum mechanics thus arises: the mass of a QCD bound-state is the sum of the mass-scales characteristic of the constituents plus a (negative and sometimes large) binding energy.

Since QCD's interactions are universal, similar cancellations must take place within the proton. However, in the proton channel, no symmetry requires the cancellations to be complete. Hence, the proton's mass has a value that is typical of the magnitude of scale breaking in QCD's one body sectors, \emph{viz}.\ the dressed-gluon and -quark mass-scales.

\section{Empirical Consequences of Emergent Mass}
The perspective described above may be called the ``DCSB paradigm''.  It provides a basis for understanding why the mass-scale for strong interactions is vastly different to that of electromagnetism, why the proton mass expresses that scale, and why the pion is nevertheless unnaturally light.  In this picture, no significant mass-scale is possible in QCD unless one of commensurate size is expressed in the dressed-propagators of gluons and quarks.  Consequently, the mechanism(s) responsible for the emergence of mass can be exposed by measurements sensitive to such dressing.  Herein, three examples will be described.

Regarding the pion's valence-quark parton distribution function (PDF), one of the earliest predictions of the QCD parton model is \cite{Ezawa:1974wm, Farrar:1975yb, Berger:1979du}:
\begin{equation}
\label{PDFQCD}
{\mathpzc u}^{\pi}(x;\zeta =\zeta_H) \sim (1-x)^{2}\,,
\end{equation}
where $\zeta_H$ is an energy scale characteristic of nonperturbative dynamics.  The exponent evolves as $\zeta$ increases beyond $\zeta_H$, becoming $2+\gamma$, where the anomalous dimension $\gamma\gtrsim 0$ increases as $\ln \zeta$.  $\pi N$ Drell-Yan measurements are well suited to extracting ${\mathpzc u}^\pi(x;\zeta)$; but existing measurements are thirty years old \cite{Badier:1980jq, Badier:1983mj, Betev:1985pg, Falciano:1986wk, Guanziroli:1987rp, Conway:1989fs}; and conclusions drawn from those experiments are controversial \cite{Holt:2010vj}.
Using a leading-order (LO) analysis of their data, Ref.\,\cite{Conway:1989fs} (E615 experiment) reported ($\zeta_5 = 5.2\,$GeV):
${\mathpzc u}_{\rm E615}^{\pi}(x; \zeta_5) \sim (1-x)^{1}$,
a striking contradiction of Eq.\,\eqref{PDFQCD}.  Subsequent calculations \cite{Hecht:2000xa} confirmed Eq.\,\eqref{PDFQCD}, prompting reconsideration of the E615 analysis, with the result that, at next-to-leading order (NLO) and including soft-gluon resummation \cite{Wijesooriya:2005ir, Aicher:2010cb}, the E615 data can be viewed as consistent with Eq.\,\eqref{PDFQCD}.
Notwithstanding these advances, uncertainty remains because more recent  analyses of the E615 data have failed to incorporate threshold resummation effects and, crucially, modern data are lacking.

\begin{figure}[t]
\leftline{%
\includegraphics[clip, width=0.44\textwidth]{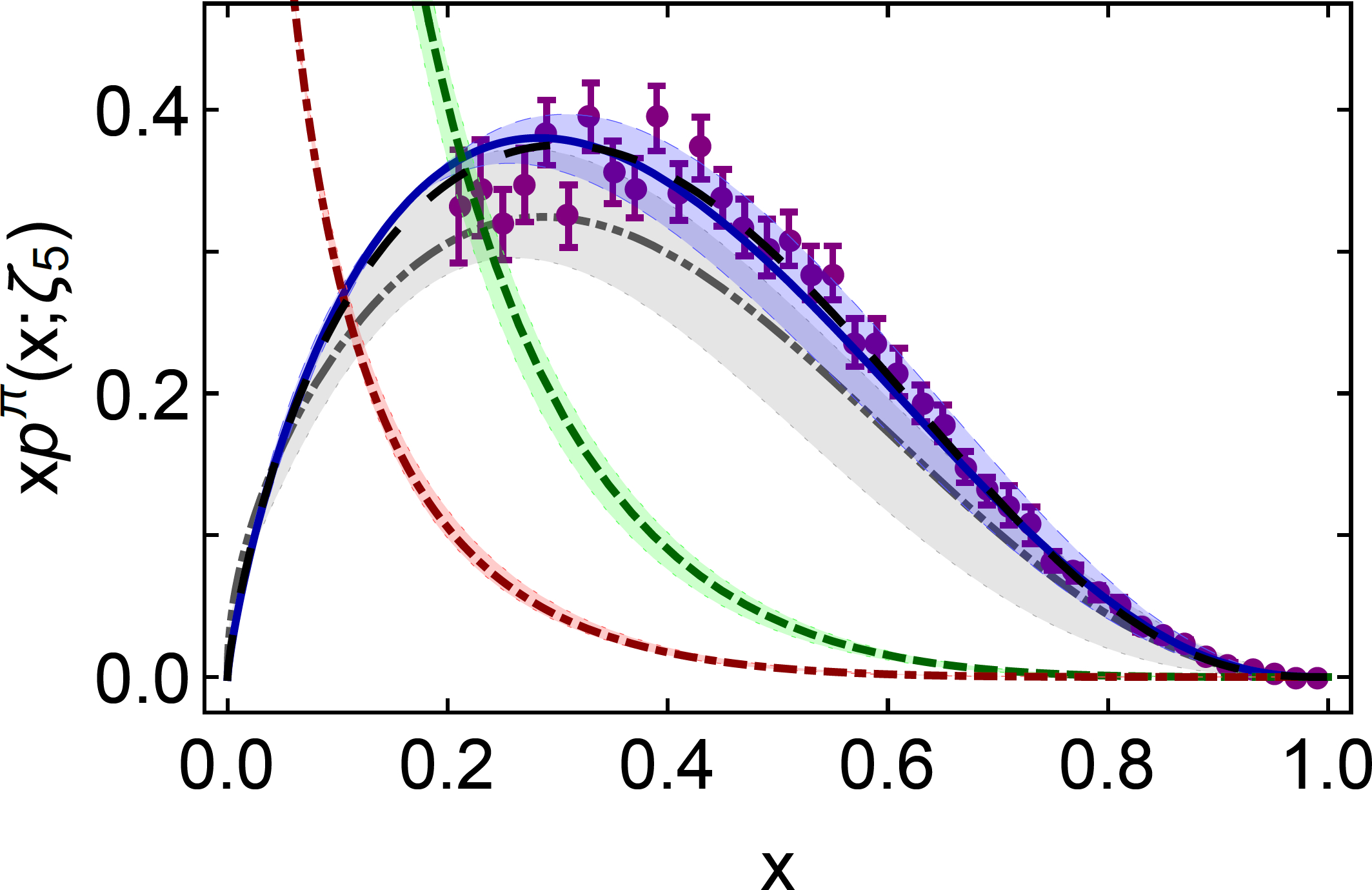}}
\vspace*{-34ex}

\hspace*{0.5\textwidth}\parbox[t]{0.44\textwidth}{\caption{\label{figF12} \small
Pion valence-quark momentum distribution function, $x {\mathpzc u}^\pi(x;\zeta_5)$:
dot-dot-dashed [grey] curve -- lQCD result \cite{Sufian:2019bol};
long-dashed [black] curve -- early continuum analysis \cite{Hecht:2000xa};
and solid [blue] curve  -- modern continuum calculation \cite{Ding:2019lwe}.
Gluon momentum distribution, $x g^\pi(x;\zeta_5)$ -- dashed [green] curve;
and sea-quark momentum distribution, $x S^\pi(x;\zeta_5)$ -- dot-dashed [red] curve.
(The shaded bands indicate the size of calculation-specific uncertainties \cite{Ding:2019lwe}.)
Data [purple] from Ref.\,\cite{Conway:1989fs}, rescaled following Ref.\,\cite{Aicher:2010cb}.}}
\end{figure}

Pressure is now being applied by modern advances in theory.  Lattice-QCD algorithms are beginning to yield results for the pointwise behavior of ${\mathpzc u}^\pi(x;\zeta)$ \cite{Sufian:2019bol}.
Moreover, extensions of the continuum analysis in Ref.\,\cite{Hecht:2000xa} have yielded the first parameter-free predictions of the valence, glue and sea distributions within the pion \cite{Ding:2019lwe}; and revealed that, like the pion's parton distribution amplitude (PDA), the valence-quark distribution function is hardened by DCSB, \emph{i.e}.\ as a direct consequence of emergent mass.
Significantly, as evident in Fig.\,\ref{figF12}, the new continuum prediction for ${\mathpzc u}^\pi(x;\zeta_5)$ matches that obtained using lQCD \cite{Sufian:2019bol}.  A modern confluence has thus been reached, demonstrating that real strides are being made toward understanding pion structure.   The Standard Model prediction, Eq.\,\eqref{PDFQCD} is stronger than ever before; and an era is dawning in which the ultimate experimental checks can be made \cite{R.A.Montgomery:2017hab, Denisov:2018unj, Aguilar:2019teb}.

\begin{figure}[t]
\begin{tabular}{lr}
\vspace*{3ex}
\includegraphics[clip, width=0.44\textwidth]{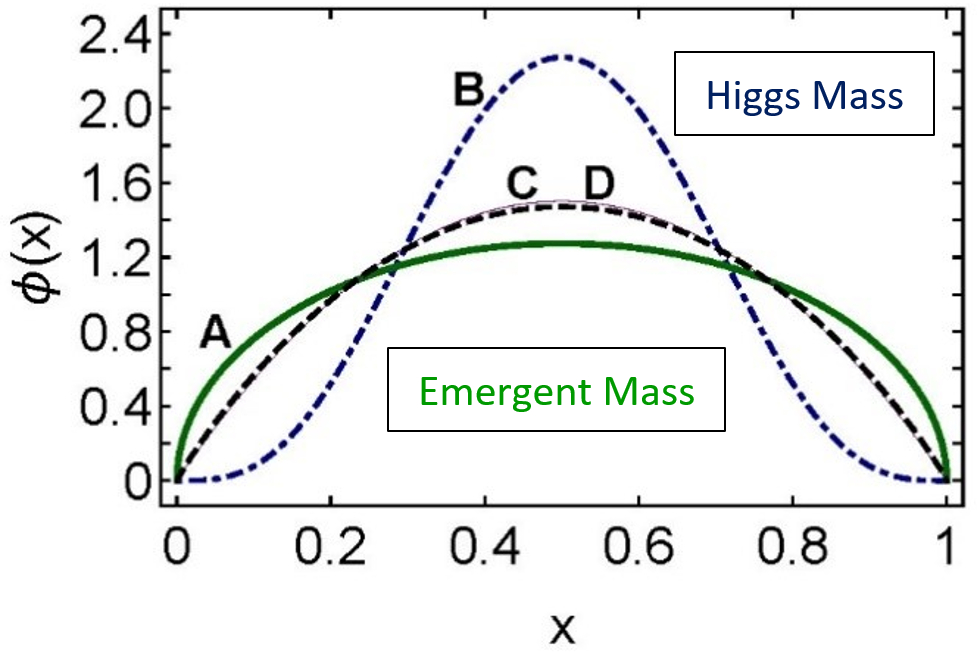} \hspace*{2ex} &
\hspace*{2ex}\includegraphics[clip, width=0.45\textwidth]{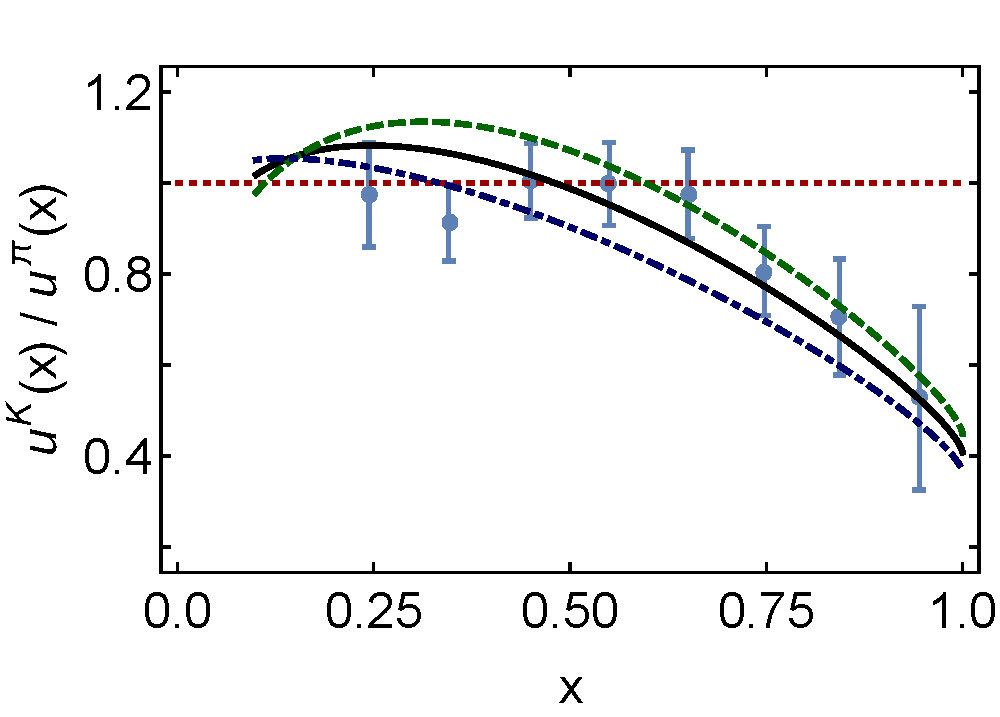}
\end{tabular}
\caption{\label{figPDAs}\small
\underline{Left Panel}.
Twist-two parton distribution amplitudes at a resolving scale $\zeta=2 \,$GeV$=:\zeta_2$. \textbf{A} solid (green) curve – pion $\Leftarrow$ emergent mass generation is dominant; \textbf{B} dot-dashed (blue) curve – $\eta_c$ meson $\Leftarrow$ Higgs mechanism is the primary source of mass generation;  \textbf{C} solid (thin, purple) curve -- asymptotic profile, 6x(1 - x); and \textbf{D} dashed (black) curve – ``heavy-pion'', \emph{i.e}.\ a pion-like pseudo-scalar meson in which the valence-quark current masses take values corresponding to a strange quark $\Leftarrow$ the boundary, where emergent and Higgs-driven mass generation are equally important.
\underline{Right Panel}.
Ratio of valence $u$-quark PDFs in the pion and the kaon at $\zeta = 5.2\,$GeV$=:\zeta_5$.
Data are from Drell-Yan measurements \cite{Badier:1980jq}.
The computed results are taken from Ref.\,\cite{Chen:2016sno}, with the dashed, solid, and dot-dashed curves representing, respectively, $0$, $5$\%, $10$\% of the kaon's light-front momentum carried by glue at the scale, $\zeta_K = 0.51\,$GeV.
%
%
}
\end{figure}

The potential of such measurements to expose emergent mass is greatly enhanced if one includes similar kaon measurements.  This can be illustrated by considering a class of meson ``wave functions'', a number of which are depicted in Fig.\,\ref{figPDAs}\,--\,Left Panel.  This image answers the following question: When does the Higgs mechanism begin to influence mass generation?  In the limit of infinitely-heavy quark masses; namely, when the Higgs mechanism has overwhelmed every other mass generating force, the PDA becomes $\delta(x-1/2)$. The heavy $\eta_c$ meson, constituted from a valence charm-quark and its antimatter partner, feels the Higgs mechanism strongly. On the other hand, contemporary continuum- and lQCD calculations predict that the PDA for the light-quark pion is a broad, concave function \cite{Chang:2013pq, Zhang:2017bzy}. Such features are a definitive signal that pion properties express emergent mass generation. The remaining example in Fig.\,\ref{figPDAs} shows that the PDA for a system composed of strange quarks almost matches that of QCD’s asymptotic (scale-free) limit.  Hence, this system lies at the boundary, with strong (emergent) mass generation and the weak (Higgs-connected) mass playing a roughly equal role.

Consequently, comparisons between distributions of truly light quarks and those describing strange quarks are ideally suited to exposing measurable signals of emergent mass in counterpoint to Higgs-driven effects; and a most striking example can be found in the contrast between the valence-quark PDFs of the pion and kaon at large Bjorken-$x$.  A significant disparity between these distributions would point to a marked difference between the fractions of pion and kaon momentum carried by the other bound state participants, particularly gluons.

A prediction for the ratio ${\mathpzc u}^K(x)/{\mathpzc u}^\pi(x)$ is available \cite{Chen:2016sno}.  Reproduced in Fig.\,\ref{figPDAs}\,--\,Right Panel, the result confirms this assessment: agreement with data \cite{Badier:1980jq} indicates that the gluon content of the kaon at $\zeta_K=0.51\,$GeV is just $5\pm 5$\%, whereas that for the pion is more-than 30\%.  Hence, there are striking differences between the gluon content of the pion and kaon; and they persist to large resolving scales, \emph{e.g}.\ at $\zeta = 2\,GeV $ the gluon momentum fraction in the pion is still 50\% greater than that in the kaon. This difference in gluon content is clearly expressed in the large-$x$ behaviour of the $\pi$ and $K$ valence-quark PDFs.  It is a striking empirical signal of the almost pure NG-boson character of the pion, marking the near perfect expression of Eq.\,\eqref{EasyOne} in this almost-massless state as compared to the incomplete cancellation in the $s$-quark-containing kaon.  The issue again, however, is that there is only one forty-year-old measurement of ${\mathpzc u}^K(x)/{\mathpzc u}^\pi(x)$ \cite{Badier:1980jq}.

\section{Epilogue}
No claim to have understood the Standard Model is supportable until an explanation is provided for the emergence and structure of Nambu-Goldstone (NG) modes in the Standard Model.  NG modes are far more complex than is typically thought.  They are not pointlike; they are intimately connected with the origin of mass; and they probably play an essential part in any answer to the question of gluon and quark confinement in the physical Universe.  The internal structure of NG modes is complex; and that structure provides the clearest window onto the emergence of mass in the Standard Model.  The cleanest expression of this is found in the statement that the gluon content within Nature's only near-pure Nambu-Goldstone mode, the pion, is far greater than that in any other hadron.  This is observably expressed in the pion's valence-quark distribution function and very strikingly in a comparison between the valence-quark distributions in the pion and much-heavier kaon.  New-era experiments capable of validating these predictions are therefore of the highest priority.  With validation, an entire chapter of the Standard Model, whose writing began more than eighty years ago \cite{Yukawa:1935xg}, can be completed and closed with elucidation of the structural details of the Standard Model's only NG modes, whose existence and properties are critical to the formation of everything from nucleons, to nuclei, and on to neutron stars.

The LHC has not found the ``God particle'' because the Higgs boson is not the origin of mass.  Concerning everyday nuclei, the Higgs mechanism only produces a little bit of mass.  It explains neither the large value of the proton's mass nor the pion's (near-)masslessness.  The key to understanding the origin and properties of the vast bulk of all known matter is the strong interaction sector of the Standard Model.  The current paradigm is QCD, plausibly the only mathematically well defined four-dimensional theory that science has ever produced.  Hence, the goal is to reveal the content of strong-QCD.  In working toward this, no one approach is sufficient.  Progress and insights are being delivered by an amalgam of experiment, phenomenology and theory; and continued exploitation of the synergies between these efforts is essential if nuclear physics is to capitalise on new opportunities provided by existing and planned facilities.

\ack

This discussion is based on work completed by an international collaboration involving many remarkable people, to all of whom I am greatly indebted.  It is complemented by other contributions in this volume, \emph{e.g}.\ those from R.\,A.~Montgomery and J.~Rodr{\'{\i}}guez-Quintero.
I would like to express my gratitude to the organisers of the 27th International Nuclear Physics Conference (INPC 2019), who ensured that the meeting was a success and my participation was both enjoyable and fruitful.
Work partly supported by
Jiangsu Province \emph{Hundred Talents Plan for Professionals}.
%


\providecommand{\newblock}{}

\end{document}